\documentclass[conference]{IEEEtran}
\IEEEoverridecommandlockouts

\usepackage{cite}
\usepackage{amsmath,amssymb,amsfonts}
\usepackage{algorithmic}
\usepackage{textcomp}
\usepackage{xcolor}
\usepackage{multirow}
\usepackage{wrapfig}
\usepackage[pdftex]{graphicx}

\usepackage{pifont}
\newcommand{\xmark}{\ding{55}}%

\def\BibTeX{{\rm B\kern-.05em{\sc i\kern-.025em b}\kern-.08em
    T\kern-.1667em\lower.7ex\hbox{E}\kern-.125emX}}

\begin{document}

\title{Event Camera Simulator Design for Modeling Attention-based Inference Architectures}

\author{\IEEEauthorblockN{Md Jubaer Hossain Pantho}
\IEEEauthorblockA{University of Florida \\
Gainesville, USA \\
mpantho@ufl.edu}
\and
\IEEEauthorblockN{Joel Mandebi Mbongue}
\IEEEauthorblockA{University of Florida \\
Gainesville, USA \\
jmandebimbongue@ufl.edu}
\and
\IEEEauthorblockN{Pankaj Bhowmik}
\IEEEauthorblockA{University of Florida \\
Gainesville, USA \\
pankajbhowmik@ufl.edu}
\and
\IEEEauthorblockN{Christophe Bobda}
\IEEEauthorblockA{University of Florida \\
Gainesville, USA \\
cbobda@ece.ufl.edu}
}

\maketitle

\begin{abstract}
In recent years, there has been a growing interest in realizing methodologies to integrate more and more computation at the level of the image sensor. The rising trend has seen an increased research interest in developing novel event cameras that can facilitate CNN computation directly in the sensor. However, event-based cameras are not generally available in the market, limiting performance exploration on high-level models and algorithms. This paper presents an event camera simulator that can be a potent tool for hardware design prototyping, parameter optimization, attention-based innovative algorithm development, and benchmarking. The proposed simulator implements a distributed computation model to identify relevant regions in an image frame. Our simulator's relevance computation model is realized as a collection of modules and performs computations in parallel. The distributed computation model is configurable, making it highly useful for design space exploration. The Rendering engine of the simulator samples frame-regions only when there is a new event. The simulator closely emulates an image processing pipeline similar to that of physical cameras. Our experimental results show that the simulator can effectively emulate event vision with low overheads.
\end{abstract}


\section{Introduction}
Event cameras are bio-inspired vision sensors designed to generate image frames asynchronously based on scenic events \cite{eventSurvey}. In contrast to conventional camera sensors where raw frame pixels are streamed to a backend processor at a fixed rate, event-based cameras generate output only when there is a new event(s). Recently, researchers are seeking novel methodologies to incorporate machine learning models (in particular CNNs) in the image sensor \cite{cannici2019}\cite{Pantho2021}. This has revived interests in event cameras to facilitate efficient dataflow between the sensor and the near-sensor processing system. However, novel algorithms and methods are required to process the unorthodox data streams from these vision sensors to unlock their full potential \cite{Gehrig2018}. However, researchers working on this domain face two major challenges. \textbf{First}, there are not sufficient event-cameras in the market, limiting the research to a few applications. \textbf{Second}, the commercially available event cameras suffer from different setbacks such as low resolution, lack of reconfiguration, etc.

Several camera simulators have been proposed in the literature to accommodate the research demands \cite{esim2018}\cite{Reichel2015}. For instance, authors in \cite{esim2018}, presented ESIM, a camera simulator that resembles an event camera's behavior. The simulator integrates an adaptive rendering scheme that only samples frames when necessary. In addition to generating events, the simulator can produce a depth map, motion field, and camera trajectory. However, the simulator was developed for robotics applications and not specifically designed to explore inference architectures near the sensors. Therefore,  any in-sensor high-level processing engine that aims to leverage the event sensor in the processing pipeline will fail to utilize the full potential of the events generated from this camera simulator. At best, the simulator would allow the inference engine to only activate whenever a new event is detected on the sensor interface. However, at each iteration, the full image will be processed by the inference engine regardless of the size of the ROI (Region-Of-Interest). The newest developments in imaging technology have brought forth parallel processing image sensors that can be combined with an inference engine to provide high-performance computation models near the sensor \cite{eventSurvey}\cite{Chen2020}\cite{sakakibara2018}. By tightly coupling computation on the inference layer to specific image regions, it is possible to improve the computational capabilities of these systems and reduce data communications. Nevertheless, a suitable platform is required to explore the design space of these architectures.

\begin{figure}
    \centering
    \includegraphics[width = \linewidth]{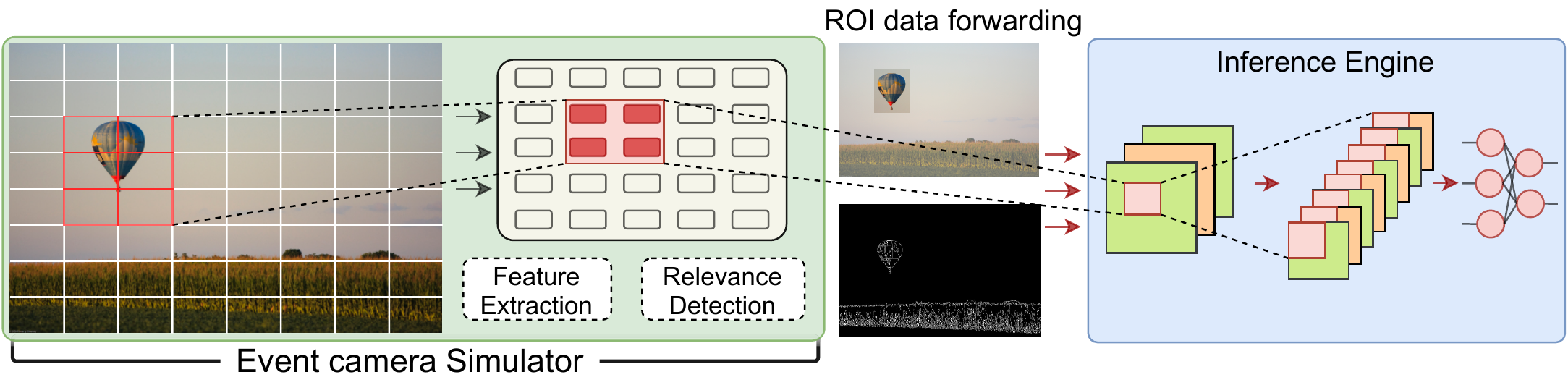}
    \caption{Region-based event camera simulator designed to accommodate inference processing near the sensor}
    \label{fig:motivation}
\end{figure}

In this paper, we present a novel event camera simulator that simulates a per-pixel image sensor's behavior aiming to accommodate CNN inference in the sensor interface. The events captured in the simulator are identified on a region-level. Therefore only specific regions can be forwarded to the following computation layer to activate the inference engine minimally (shown in figure \ref{fig:motivation}). Similar to the work mentioned above, our rendering-module samples image frames whenever there is a new event. However, instead of sampling the complete image, respective event regions are only sampled. The simulator can generate valid event data from a video stream that can be used to model and train event-based learning models. We have prototyped the simulator's computation module on an FPGA to estimate the hardware cost. Our evaluation results suggest that we can significantly reduce computation with our event-based camera approach with minimum hardware overhead.

The main contributions of this paper are:

\begin{itemize}
    \item A novel camera simulator design that identifies events on a region-basis and facilitate suitable interface for inference architectures.
    \item A thorough evaluation of our region-level relevance computation model to highlight significance.
    \item An FPGA prototype of the relevance computation model to indicate hardware overheads related to our approach. 
\end{itemize}

The remaining sections of this paper are organized as follows. Section \ref{sec:related} discusses the related works in the literature. Section \ref{sec:proposed} provides a detailed explanation of our design. We evaluate our model in Section \ref{sec:results}.
\section{Related Work}\label{sec:related}
Several camera simulators can be found in the literature emulating the behavior of an event camera \cite{Bi2017}\cite{pyDVS2016}\cite{Katz2012}. And, in recent years, various approaches have been proposed to bring inference computation close to the sensor. We start by studying the state-of-the-art camera simulator and highlight the advantages of our proposed toolchain. Next, we will discuss the in-sensor processing architectures that leverage event-based camera designs.

In \cite{mueggler2017}, authors present an event sensor simulator that can render events from a 3D scene. The simulator was designed to facilitate research in robotic vision. However, it is not tailored for in-sensor processing exploration. The virtual camera proposed in \cite{InteriorNet18} offers an interactive interface with a custom rendering engine that can be used for benchmarking different SLAM algorithms. Similar to previous work, here, the authors did not illustrate the use cases with inference architectures but focused on generating photo-realistic indoor scenes datasets. 

We found ESIM as one of the thorough works on event camera simulation \cite{esim2018}. It provides an open-source design and illustrates use cases on learning optical flow. However, ESIM (including all the other works described above) identifies events at a pixel level. These fine-grained events captured in the sensor interface can reduce the rendering engine's workload; nevertheless, the subsequent CNN accelerator in the processing pipeline fails to leverage much benefit from these fine-grained events due to the available dataflow mechanisms.

The ReImagine program launched by DARPA aims to integrate revolutionary capabilities in the imaging system \cite{Whitney2020}. They demonstrated that a single, reconfigurable ROIC (ReadOut Integrated Circuit) architecture could accommodate multiple modes of imaging operations that may be defined after a chip has been designed. The program seeks ROI-based efficient computation models to enable real-time analysis. Even though preliminary works have shown promising results, the landscape of the high-level computation part is still in progress. Further development in this research direction faces setbacks due to the lack of appropriate physical cameras that can accommodate these operations.

Other works in accommodating CNNs in an image sensor involve coupling an array of pixel processors to a parallel processing camera \cite{pantho2020}\cite{scamp52018}. Authors in \cite{pantho2020} proposed a region-aware processing model to reduce high-level computation to relevant regions. However, the authors mainly discussed the hardware aspects of the architecture. Whereas it is essential to thoroughly assess the behavior of region-aware processing models for different applications. For instance, the methodologies and threshold values used to identify relevant image regions can differ for different scenarios.

 Our simulator design differs from the works mentioned above by considering the CNN computation models that will operate on the sensor's collected data. The approaches found in the literature provide solutions at best for generic use cases and do not consider the high-level computation part. The simulator emulates event cameras that capture changes at a regional level as opposed to pixel-level sampling. This allows the subsequent computation layers to skip computation on irrelevant regions. We believe our simulators will enable researchers to develop optimized attention-based hardware architectures by accurately analyzing the relevance model. Besides, the configurability of the simulator allows exploration of the design space for event cameras.

\section{Proposed Design Flow}\label{sec:proposed}
In this section, we first describe the concept and the principles of operation of the event camera that we simulate. Then, we illustrate the design flow and architecture of the simulator.  

\subsection{Camera Model}
Our virtual camera's baseline design considers a parallel imager, where each sensing unit in the photodiode array has an analog to digital converter (ADC) and a local memory \cite{sakakibara2018}. At the sensor interface, the incoming image frame is logically divided into M image regions where $N \times N$ pixels reside in each image patch (shown in figure \ref{fig:motivation}). There is a regional processing unit (RPU) for each image patch for the local handling of computation. Each RPU has one streaming channel to transfer pixel/event data from the corresponding region to the next buffer (or computation module). All RPUs operate independently and generate output in parallel. Within the RPU, the saliency data for the corresponding region is computed. A saliency score is calculated to reflect the spatial and temporal relevance of that region. Based on the saliency score, only specific image regions are forwarded to the next plane to enable attention-based near-sensor computation.

\begin{figure}
    \centering
    \includegraphics[width = \linewidth]{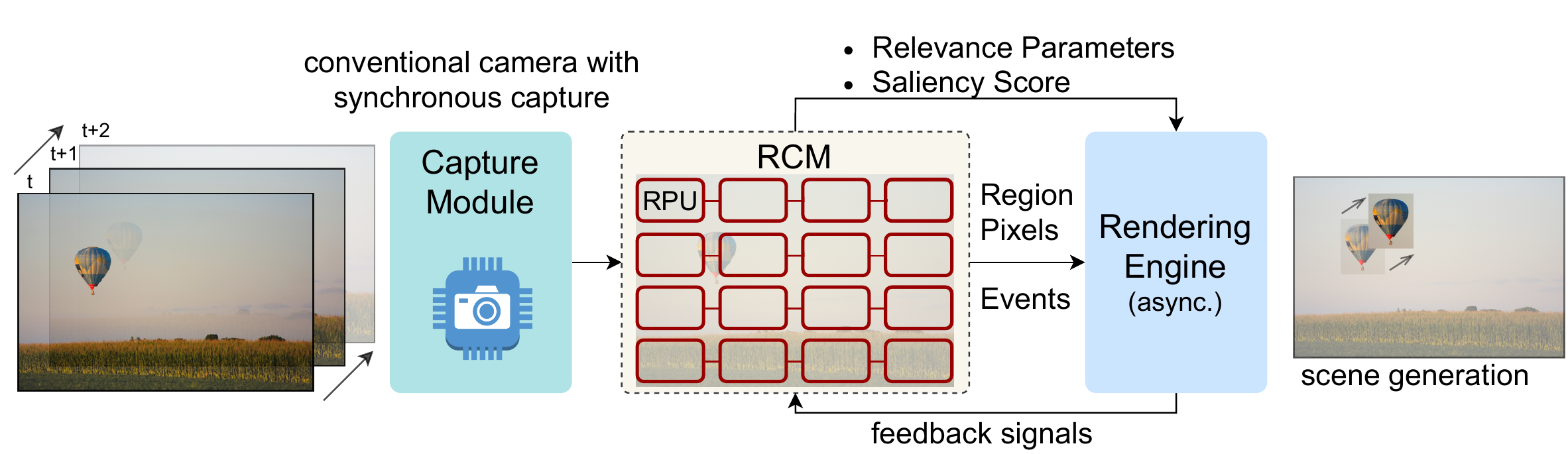}
    \caption{Proposed Simulator model.}
    \label{fig:proposed}
\end{figure}

\subsection{Simulator Architecture}
The difference between a conventional camera and an event camera is the latter does not capture intensity information from the scene synchronously. Instead, it samples visual signals asynchronously and independently for each pixel/region. In our design, we simulate this behavior with a regular vision system. The simulator's input is a stream of image frames from a camera or video clip captured at discrete time intervals. Whereas the output of the simulator includes localized pixel and event information generated at irregular intervals. The simulator comprises a capture module, a relevance computation module(RCM), and a rendering module. The high-level simulator architecture is shown in figure \ref{fig:proposed}. The capture module collects image frames at a regular interval, divides the image frame into equal-sized image patches, and forwards them to the RCM. The RCM comprises an array of RPUs operating in parallel. Within the RPU, saliency scores are computed. The saliency scores are calculated based on spatial and temporal information. Visual attention can be drawn from different details embedded within the image pixels (i.e., edges, corners, motion, error surface, optical flow, data distribution). If the saliency score is greater than some threshold, then that region is identified as relevant. The renderer collects data from the RCM and constructs the image frame for the high-level processing units in the image processing pipeline. This includes raw pixel data, saliency score, and other feature information calculated to identify the region of interest (ROI). The rendering engine renders an image at time $t$ based on the events captured at time $t$ interval and the renderer's previous state at time $t-1$. Therefore, if we denote the renderer output as $R$, it can be written as:

\begin{equation}\label{eq:render}
    R(t) = R_{ROI_{spatial}}(t-1) + R_{ROI_{temporal}}(t) 
\end{equation}

The next section details the relevance computation model utilized in our simulator.

\subsection{Relevance Computation Module (RCM)}
\begin{figure}
    \centering
    \includegraphics[width = \linewidth]{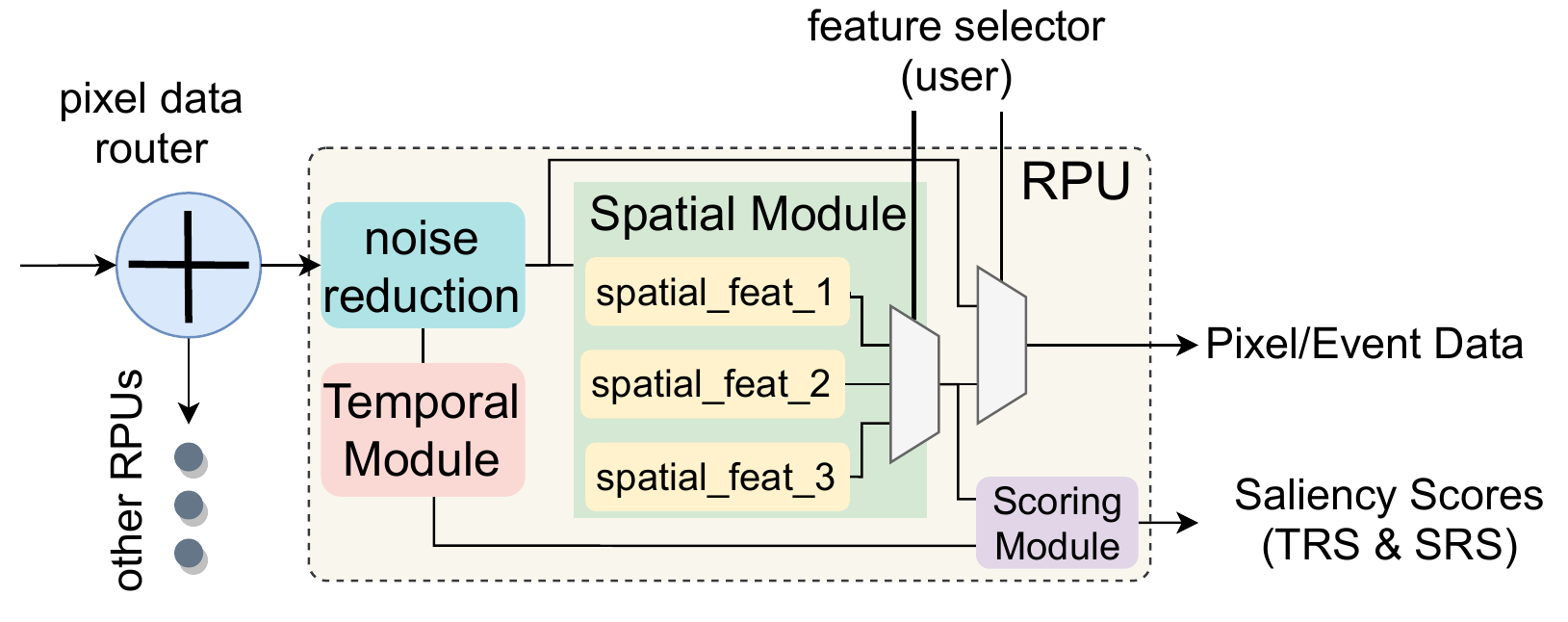}
    \caption{RPU block diagram.Here, spatial\_feat\_i indicates feature indexes used to identify spatially relevant regions (i.e. edges, corners, optical flow, etc.).}
    \label{fig:rpu}
\end{figure}

An image processing pipeline with a vision sensor and a high-level back-end processor imitates the eye and brain's combined functionality. Except, a human eye has around 130 million pixels, with only 1.3 million synaptic 170 connections to the brain, indicating a 1\% sparsity \cite{Whitney2020}. It is believed that the massive sparsity is essential for power and latency trade space and helps avoid sending repetitive information to the latter parts of the brain. The RCM of our simulator is designed to emulate the behavior of the biological vision system. This means that the RCM will receive a large number of incoming pixels from the sensor interface and forward a limited number of pixels from specific image regions to the higher processing module. The RPUs in this module operate on a region parallel basis. The RPU performs the relevance function on image pixels and accumulates the relevance score for all pixels in a region. The spatial relevance score can be calculated from a set of indexes based on the user-defined environment (i.e., edge, corners, variance, segmentation, etc.). For instance, if we consider edge points as spatial relevance index, we count the number of edge points found in an image region. Then, we use this value to rank the image regions based on a predefined threshold. Likewise, to check the spatial data distribution, the RPU can calculate the mean absolute deviation and classify the image regions based on data variation in a similar manner. Our proposed simulator implements a number of spatial relevance detection functions, from where the user can select the appropriate method that best suits a given scenario/dataset. The functionality of the RPU is shown in figure \ref{fig:rpu}. Here, the noise reduction module is used to remove noise and interference from the incoming image region. It helps to reduce the miss-detection of events.

For temporal saliency, RPUs compare the incoming pixel to its temporal neighbors. The number of temporal mismatches within a region is compared against a temporal threshold value to determine temporal relevance. The image patches are categorized using two-bit information, each for spatial and temporal saliency. This information is forwarded to the rendering engine that requests data from the RCM module based on the relevance score. The operation of the rendering engine based on the relevance score is shown in Table \ref{tab:saliencescore}. 

\begin{table}[hbt]
\centering
\caption{Computation based on the Relevance Score}
\begin{tabular}{|c|c|c|l|} 
\hline
\textbf{TRS} &  \textbf{SRS} & \textbf{RPU} & \textbf{Rendering Engine Output} \\ 
\hline
1 & 1 & \textbf{\textit{Active}} & Driven by current state\\
\hline
0 & 1 & Inactive & Driven by previous state \\
\hline
(0/1) & 0 & Inactive & Forced to Zero/previous state\\
\hline
\end{tabular}
\label{tab:saliencescore}
\end{table}

In table \ref{tab:saliencescore}, the TRS value indicated temporal relevance score, whereas the SRS value refers to the spatial relevance score. The active notion in RPU implies that for a given input frame, new image data is forwarded to the rendering engine from that RPU.

\subsection{Pixel-level Relevance vs Region-level Relevance}
As discussed above, we identify important events in our simulator on a region-level. This indicates that we label image patches with a relevance score and not individual pixels. The approach is in contrast with popular methods where events are detected on a pixel-basis. For instance, the ESIM simulator detects events on a pixel basis and estimates based on motion, optical flow, depth, and other indexes \cite{esim2018}.

We opted for a different approach because we found that a single isolated pixel-event propagated to the subsequent processing units does not provide any high-level knowledge inference. Here, we would like to highlight that high-level knowledge is inferred with machine learning algorithms in almost all image processing pipelines. And, CNNs are the most popular among them. In CNNs, identical window-based operations are performed on each input feature point at each convolutional layer. The common approaches to carry out convolution on CNN accelerators include systolic array operations or vectored window operations. In both cases, even if we narrow down our calculation to each new eventful pixel, the dataflow mechanism will limit the accelerator's ability to maximize the performance based on the fine-grained events. In other words, the inference module will not be able to leverage the fine-grained events generated at the pixel-level. Whereas with our region-level saliency detection approach, a carefully designed inference engine can localize the computation, and any new events will initiate computation only in a specific region using a vectored window operation. Besides, It is possible to opt-out calculation on isolated pixel events residing in low-scoring image patches by adequately calibrating the event camera. We found that the pruning of redundant regions has a minimum to no impact on the accuracy of the inference model. Moreover, the approach can improve the performance of sparsity-aware models by eliminating computational redundancies from the processing pipeline. For instance, authors in \cite{Pang_2019} presented a CNN-based tiny object detection mechanism that schedules image patches to a classifier and a detector to identify objects. Here, our simulator can reduce the computation by eliminating redundant image patches early at the sensor interface. Besides, in \cite{Chen2016}, authors schedule image tiles in their accelerator architecture to perform CNN operations. The output of our simulator tags each image region with their saliency score. Therefore, by adequately eliminating low-scoring regions, our event camera model can be utilized in tile-based accelerators to improve computational efficiency.

\begin{figure}[hbt]
    \centering
    \includegraphics[width = \linewidth]{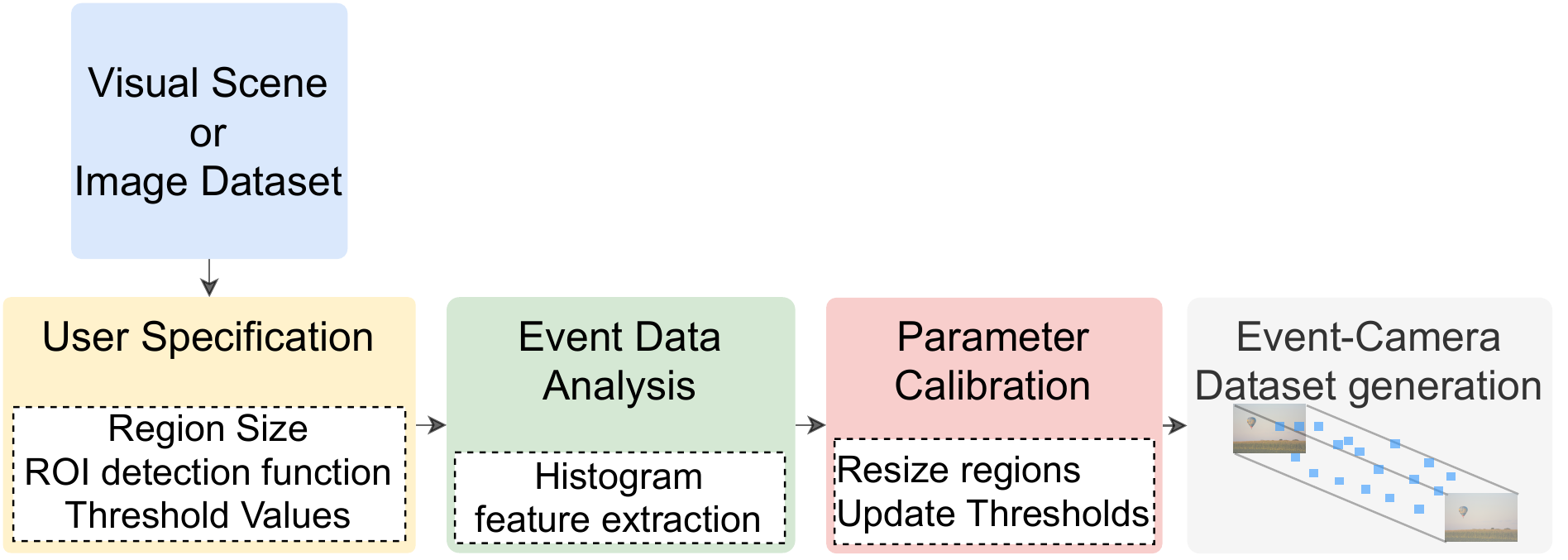}
    \caption{Simulator design flow}
    \label{fig:design-flow}
\end{figure}

\subsection{Configurability}
The benefit of our simulator is that it allows camera parameter reconfigurations for different applications. We understand that the size of the regions, the spatial relevance index, and the threshold values dictating the saliency may differ for different application scenarios. Therefore, the simulator enables users to set up these environment parameters to generate custom event-based datasets that can be later used to develop and train region-aware inference models. The design flow of our simulator is shown in figure \ref{fig:design-flow}.
\section{Results}\label{sec:results}
In this section, we detail our evaluation infrastructure and provide experimental results to indicate the efficacy of our design.

\subsection{Evaluation Infrastructure}
Our proposed simulator computes Spatio-temporal relevance to detect regions with events. However, to better evaluate the impact of the relevance function, we test the spatial and temporal modules separately for different datasets. The goal of this evaluation is to quantify the influence of our region-based relevance model. Next, we assess the effect of the region size and threshold values in our approach. Then we try to evaluate the change in accuracy for different CNN models when trained on our event-driven datasets. Finally, we prototype the RCM module on an FPGA to estimate the resource overhead of our model to evaluate the viability of realizing it at the edge. We end our evaluation by comparing our simulator with other event-based simulators found in the literature.

\subsection{Evaluation Details}
The proposed simulator is written in Python scripting language. For this evaluation, we used image datasets as the simulator's input and generated custom event-driven datasets with a reduced amount of data. For noise reduction, we used median filtering on incoming images. However, other noise reduction mechanisms can also be used. For spatial relevance detection, we implemented three feature indexes within the RPU: edge, corner, and mean absolute deviation (MAD). While edges and corners provide locality of early feature points within an image frame, the MAD value gives an insight into the statistical distribution of the region data. The edge and corner points are common feature indexes used to draw ROI in an image. Therefore, we will emphasize our evaluation of the spatial distribution of the data. Here, we chose 'mean absolute deviation' over variance due to their implementation's hardware cost. The equation for calculating variance is shown in equation \ref{eq:variance}.

\begin{equation}\label{eq:variance}
    \sigma^2 = \frac{\displaystyle\sum_{i=1}^{n}(x_i - \mu)^2} {n}
\end{equation}

Here, $\mu$ represents the mean value. Here, the square operation consumes considerable hardware resources. In contrast, MAD computation does not require square operation and has minimum hardware overhead. MAD is shown in equation \ref{eq:mad}.

\begin{equation}\label{eq:mad}
    MAD = \frac{\displaystyle\sum_{i=1}^{n}|x_i - \mu|} {n}
\end{equation}

\begin{figure}
    \centering
    \includegraphics[width = \linewidth]{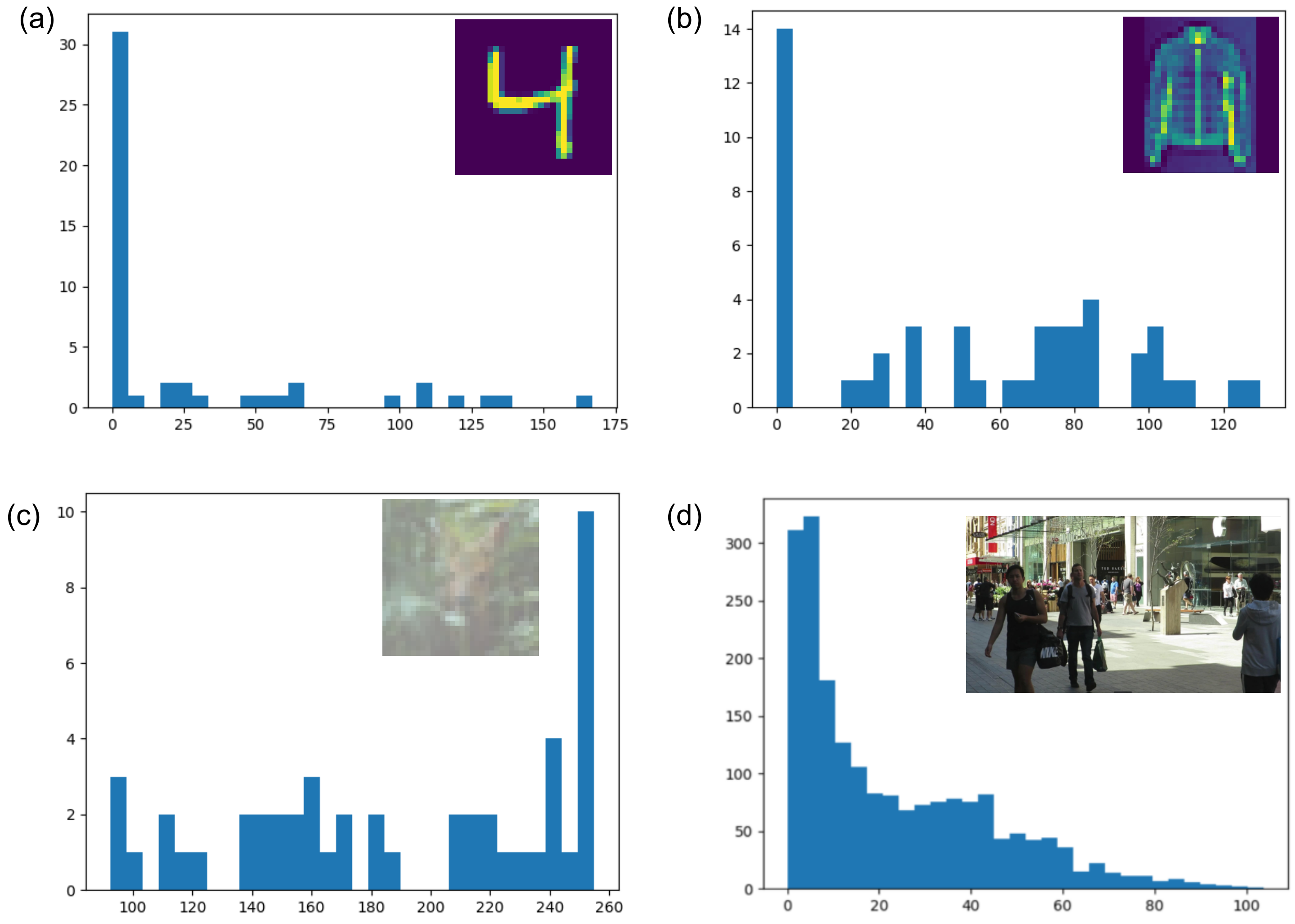}
    \caption{Distribution of Mean Absolute Deviation. For (a), (b), and (c), images are divided into 4×4 patches. In image (d), region size of 32x32 is used. (a)MNIST (b) FashionMNIST (c) CIFAR10 datasets (d) MOT17-08.}
    \label{fig:mad-data}
\end{figure}

To evaluate the effectiveness of mean absolute deviation, we first analyze the data distribution of different datasets. For this experiment, we selected four different datasets: MNIST, FashionMNIST, CIFAR10, and MOT17-08. For the first three datasets, the image size is $32\times32$, and the region size is selected to be $4\times4$. Whereas, for MOT-17 dataset, image resolution is $1920\times1080$ and we opted for a region size of $32\times32$. Figure \ref{fig:mad-data} illustrates some sample results. As we can see, for datasets (a), (b), and (d), there is a large number of regions with a MAD value close to 0. However, for image (d), this is not the case. Because in CIFAR10, the foreground to background pixel ratio is very high, and the chosen region size is comparable to the actual image size.

\subsection{Temporal Relevance Analysis}
Next, we seek to estimate the typical size of the ROI detected by the temporal module of the simulator. For this evaluation, we used the MOT17 datasets for a real-world scenario \cite{MOT16}. The dataset contains different video clips of people moving in public places. The video clips are captured with a $30$fps camera with an image resolution of $1920\times1080$. We tested our simulator on four different MOT17 datasets. Table \ref{tab:temp-roi} indicates the mean percentage of non-relevant regions for each dataset. The table indicates that more than $50\%$ of the regions contains repetitive regions over time for static camera positions.
For region-level relevance detection, it is possible to reduce a more significant amount of redundancies by carefully selecting the threshold value. Here, regions with insignificant temporal changes can be discarded from the computation. However, we notice that, for the 4th entry in the table, we have a comparatively less number of irrelevant regions due to the moving camera position. Therefore, for moving camera systems, spatial redundancy reduction techniques can be used for further improvement. The results in table \ref{tab:temp-roi} further confirm the spatiotemporal redundancy reduction technique used in our simulator.

\begin{table}[hbt]
\begin{center}
\caption{Region-level Temporal Relevance Analysis on MOT17 Datasets}
\label{tab:temp-roi}
\begin{tabular}{ |c|c|c| }
 \hline
 \textbf{Dataset} & \textbf{Description} & \textbf{Avg. ROI}  \\
 \hline
 MOT17-08 & Pedestrian street (static cam) & 41.60\% \\ 
\hline
 MOT17-03 & Sidewalk at night (static) & 25\% \\ 
 \hline
 MOT17-01 & People in a square (static) & 28.29\% \\ 
 \hline
MOT17-12 & Shopping mall (moving cam) & 69.43\% \\ 
 \hline
\end{tabular}
\end{center}
\end{table}

\begin{figure}[hbt]
    \centering
    \includegraphics[width = \linewidth]{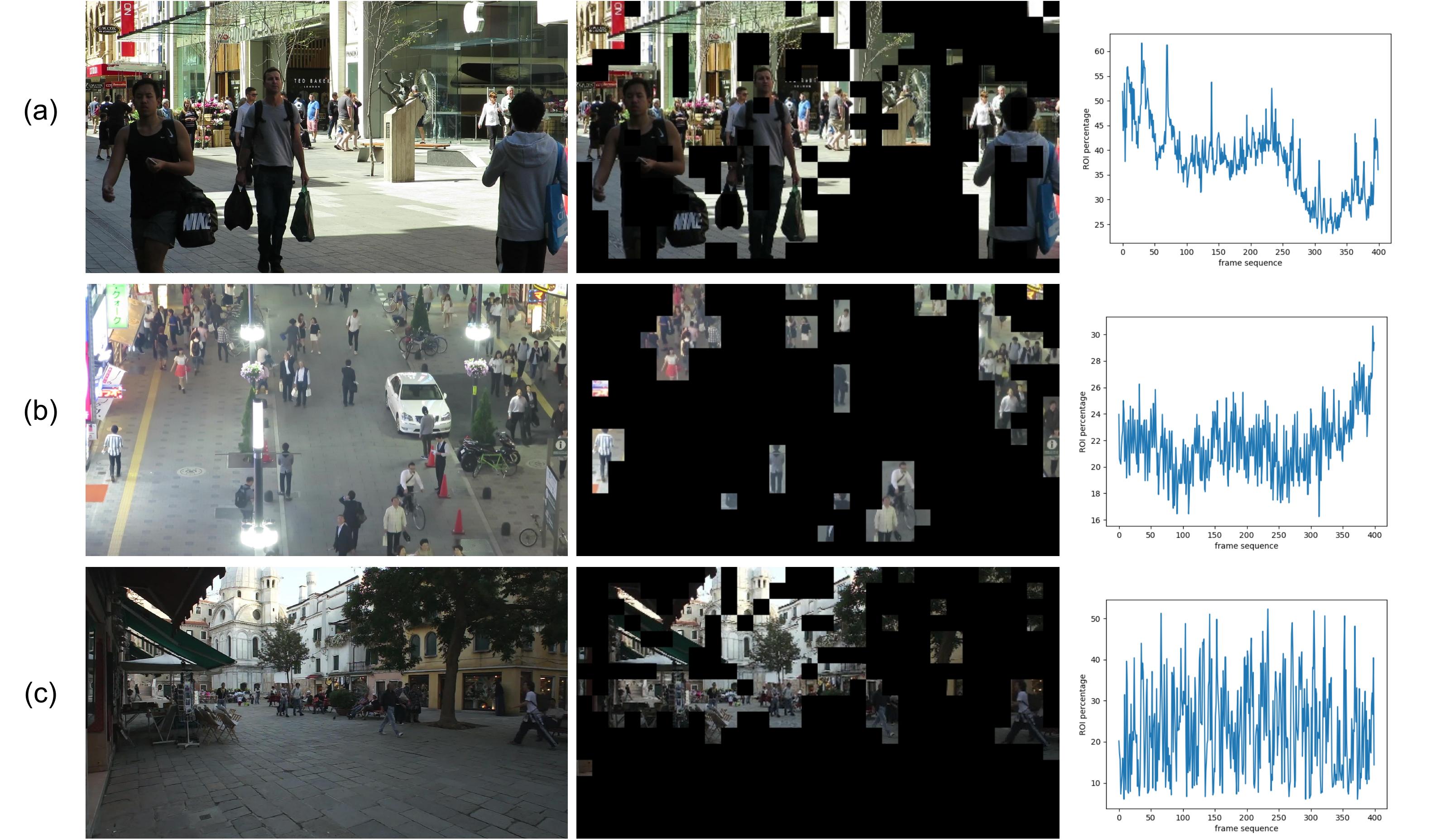}
    \caption{Region-level temporal relevance. Left column indicating original image. The second column illustrates temporal ROIs. The right column shows percentage of ROI region size over time. (a) MOT17-08 (b) MOT17-03 (c) MOT17-01}
    \label{fig:motion-pic}
\end{figure}

Figure \ref{fig:motion-pic} provides a pictorial view of the event-based outputs generated by our simulator for the MOT17 dataset. The graphs in figure \ref{fig:motion-pic} indicate the average percentage of ROI regions over time.

\subsection{Spatial Relevance Analysis}
We perform a similar study for spatial relevance detection on different datasets. We selected four datasets for this study: MNIST, FashionMNIST, OpenImages \cite{OpenImages} and mosquito species \cite{mosquito}. For the OpenImages dataset, we tested our simulator only on the airplane class due to the low foreground to background ratio on airplane images.  The average rates of spatially redundant regions in these datasets are shown in Table \ref{tab:spat-roi}. Here, the images are resized before passing them through the simulator. As the table indicates, all four datasets contain spatial redundancies that can be removed using our event-camera simulator.

\begin{table}[hbt]
\begin{center}
\caption{Region-level Spatial Relevance Analysis}
\label{tab:spat-roi}
\begin{tabular}{ |c|c|c| }
 \hline
 \textbf{Dataset} & \textbf{Image Size} & \textbf{Avg Redundancy}  \\
 \hline
 MNIST & $28\times28$ & 50\% \\ 
\hline
 FashionMNIST & $28\times28$ & 29\% \\ 
 \hline
 OpenImages (airplane) & $224\times224$ & 31\% \\ 
 \hline
 Mosquito & $224\times224$ & 40\% \\ 
 \hline
\end{tabular}
\end{center}
\end{table}

\begin{figure}
    \centering
    \includegraphics[width = \linewidth]{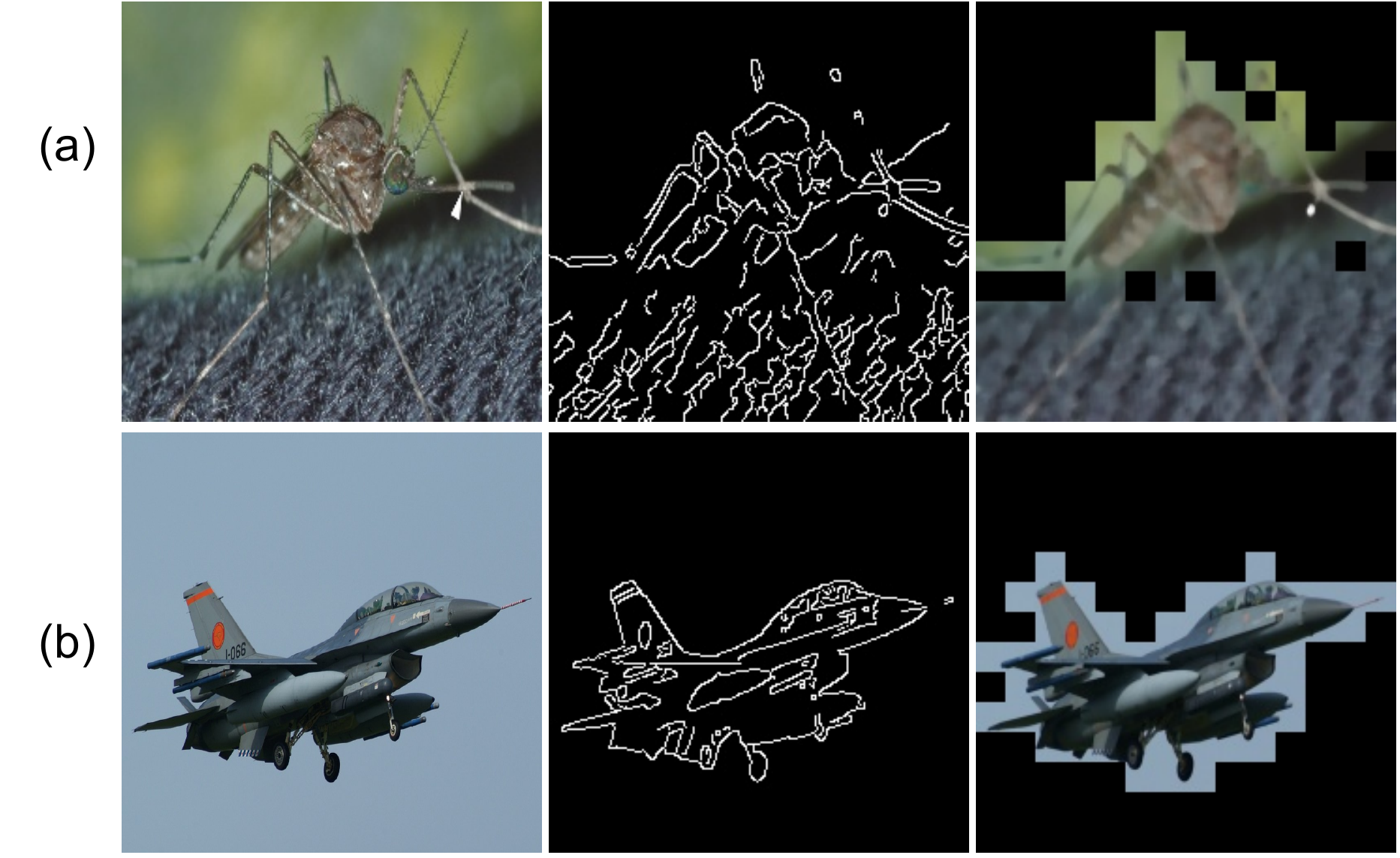}
    \caption{Region-level spatial relevance. Left column indicating original image. The second column indicates edge points as possible feature points. The right column shows output image from our simulator. (a) Mosquito \cite{mosquito} (b) OpenImages \cite{OpenImages}.}
    \label{fig:spat-pic}
\end{figure}

\begin{figure}
    \centering
    \includegraphics[width = \linewidth]{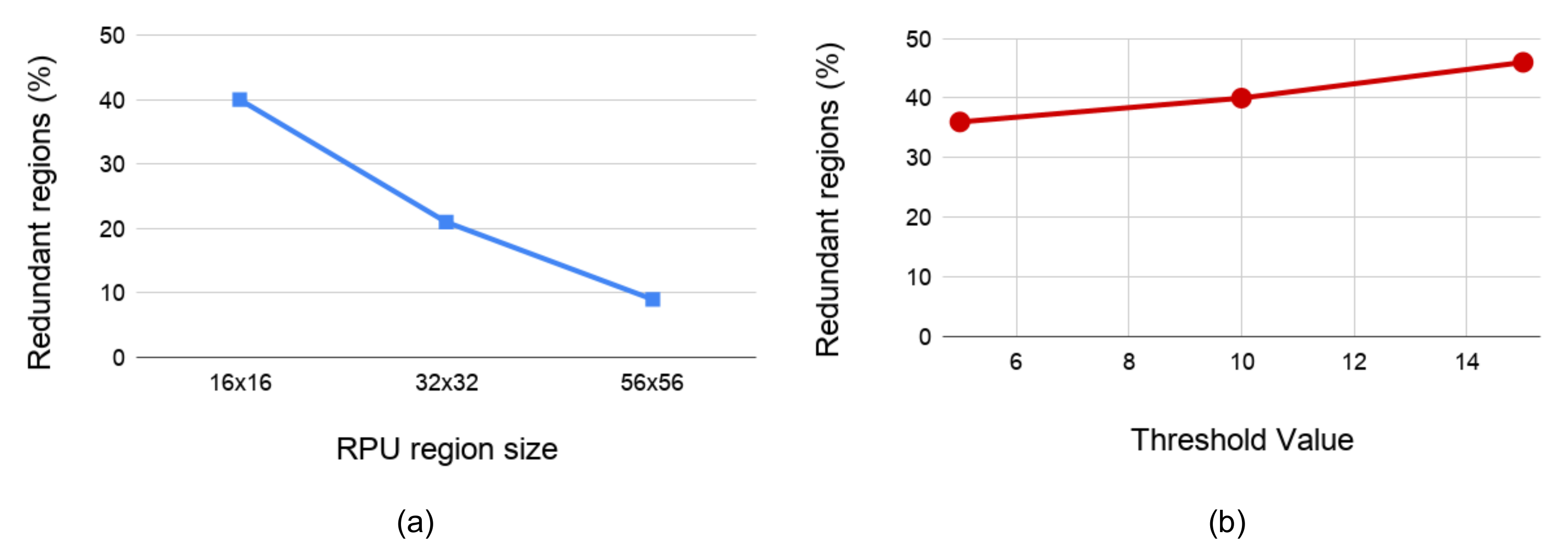}
    \caption{ Change in ROI size with (a) RPU region size and (b) threshold value (tested on Mosquito dataset \cite{mosquito}).}
    \label{fig:th-size}
\end{figure}

Figure \ref{fig:spat-pic} provides a pictorial view of the ROI detected images shown in Table \ref{tab:spat-roi}. For different datasets, region sizes are adjusted for optimal results. Here, we would like to emphasize again that the average redundancy found in these datasets is dependent on the threshold values and the region size selected for them. To better illustrate the impact, we tested different threshold values and region sizes on the Mosquito data used in table \ref{tab:spat-roi}. Figure \ref{fig:th-size} illustrates the results. Here, we calculated edge points to identify spatial redundancy. For a given threshold value, figure \ref{fig:th-size}(a) was generated. As we can see, the number of redundant regions decreases as we increase the size of the RPU region. This is because as we increase the region size, fine-grained regions get excluded from redundancy calculation. We observe a similar scenario as we decrease the threshold value. In figure \ref{fig:th-size}(b) we use a region size of $16\times16$ for calculation. However, it should be noted that increasing the threshold value too high may cause relevant regions to be incorrectly removed. Therefore, it is necessary to identify the optimal point for the threshold and RPU region size.

The spatially redundant regions are labeled with an SRS value 0, and temporally redundant regions are tagged with a TRS value of 0. Therefore, while using these datasets in CNN inference hardware such as \cite{Chen2016}, it is possible to skip computation for low SRS tiles and avoid repetitive computation for low TRS tiles.

\subsection{Impact on CNN Inference}
Next, we evaluate the impact on the accuracy of different CNN models when trained on these custom region-based ROI-extracted datasets. We tested on three different models with three different datasets. The results are listed in Table \ref{tab:accuracy}. As the table suggests, when trained on our simulator-generated datasets, we see little to no drop of inaccuracy for all the cases. However, we believe further studies can bring about even better results for even-detected datasets in the future. And our designed simulator can play a vital role in assisting these works.

\begin{table}[hbt]
\begin{center}
\caption{Impact on CNN Model Accuracy}
\label{tab:accuracy}
\begin{tabular}{ |c|c|c|c| }
 \hline
 \textbf{Dataset} & \textbf{Models} & \textbf{\vtop{\hbox{\strut Accuracy}\hbox{\strut (Original)}}} & \textbf{\vtop{\hbox{\strut Accuracy}\hbox{\strut (roi-based)}}}  \\
 \hline
 MNIST & LeNet-5 & 98.93\% & 98.8\% \\ 
\hline
 FashionMNIST & LeNet-5 & 88\%  & 87.8\% \\ 
 \hline
 Mosquito \cite{mosquito} & ResNet-50 & 99\%  & 99\% \\ 
 \hline
 Mosquito & VGG-16 & 99\% & 99\%\\ 
 \hline
\end{tabular}
\end{center}
\end{table}

\begin{wrapfigure}{r}{0.55\linewidth}
    \centering
    \includegraphics[width = \linewidth]{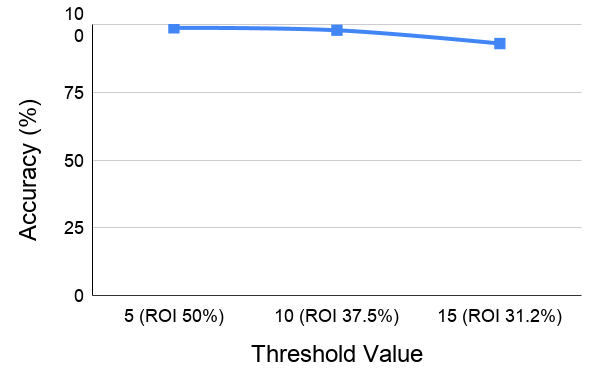}
    \caption{Change in accuracy with threshold value for MNIST data.}
    \label{fig:accuracy-vs-th}
\end{wrapfigure}
The accuracy listed in table \ref{tab:accuracy} was achieved on the simulator generated datasets with spatially redundant regions discarded (listed in table \ref{tab:spat-roi}). As we mentioned before, by increasing the threshold value, it is possible to decrease the relevant region size in images. However, it will impact the accuracy of the following CNN model as well. We tested it on the MNIST dataset for different thresholds. We see that the accuracy of the LeNet-5 model starts decreasing as we start increasing the threshold value beyond a certain point. This is shown in figure \ref{fig:accuracy-vs-th}. Here, we select the RPU region size of $8\times8$ and edge points as spatial feature index. The threshold value of 5 indicates that the number of edge points in an $8\times8$ region has to be greater than equal to 5 to be considered a relevant region.

\subsubsection{Hardware Design Evaluation}
The end goal of this research is to develop suitable inference architectures that can be integrated with a region-aware camera sensor facilitating an event-based processing pipeline at the edge. Therefore, while designing the simulator, it is necessary to adopt ROI-detection functions with minimum hardware overhead. We prototyped the RPU and the RCM module of our simulator in a Virtex UltrScale plus FPGA (VCU118) to estimate the hardware cost associated with it. We opted to realize the RCM module because this is the module that draws the visual attention in our simulator. The resource utilization is shown in Table \ref{tab:fpga}. Here, the RPU is designed for an $8\times8$ region size, and the RCM data is estimated for $224\times224$ incoming image frames. The table indicates that the RCM module only consumes \textbf{$12\%$} combinational logics available in the Virtex FPGA. This confirms the viability of its realization with available CNN acceleration engines.

\begin{table}[hbt]
\caption{FPGA Resource Utilization of the RCM}
\label{tab:fpga}
\begin{center}
\begin{tabular}{|c|c|c|c|}
\hline
Module Name & \textbf{LUT} & \textbf{FF} & \textbf{LUTRAM}\\
\hline
RPU & 183 & 90 & 16\\
\hline
RCM & 143,472 & 70,562 & 12,544\\
\hline
\end{tabular}
\end{center}
\end{table}

Finally, we perform a qualitative comparison to our work with existing camera simulators found in the literature. This is shown in table \ref{tab:comp}.

\begin{table}[hbt]
\caption{Simulator Design Comparison}
\label{tab:comp}
\begin{center}
\begin{tabular}{|c|c|c|c|}
\hline
- & \textbf{\cite{esim2018}} & \textbf{\cite{Kaiser2016}} & \textbf{Ours}\\
\hline
 \vtop{\hbox{\strut Transmit Events}\hbox{\strut along with frame}} & \checkmark & \checkmark & \checkmark\\
\hline 
Adaptive Rendering & \checkmark & \xmark & \checkmark\\
\hline
Events detected & pixel & pixel & region\\
\hline
Configurability & \checkmark & N/A & \checkmark \\
\hline
\end{tabular}
\end{center}
\end{table}

\section{Conclusion}\label{sec:Conclusion}
This paper presents an event-camera simulator that emulates the behavior of an attention-based parallel camera sensor. The simulator computes the relevant score for each region and performs rendering operations for only relevant regions. The region-based ROI detection model adopted in this work can provide high-performance computing for high-level reasoning models. Our proposed simulator will serve as an analyzing tool to develop machine learning models that can best explore the event camera in the processing chain. The ROI detecting functions used in the simulator have low hardware cost. This makes it viable to implement in a distributed architecture. Our experimental results show that the attention-based approach used in this work can significantly reduce operation execution for inference engines.

\bibliographystyle{IEEEtran}
\bibliography{IEEEabrv, main}

\begin{thebibliography}{10}
\providecommand{\url}[1]{#1}
\csname url@samestyle\endcsname
\providecommand{\newblock}{\relax}
\providecommand{\bibinfo}[2]{#2}
\providecommand{\BIBentrySTDinterwordspacing}{\spaceskip=0pt\relax}
\providecommand{\BIBentryALTinterwordstretchfactor}{4}
\providecommand{\BIBentryALTinterwordspacing}{\spaceskip=\fontdimen2\font plus
\BIBentryALTinterwordstretchfactor\fontdimen3\font minus
  \fontdimen4\font\relax}
\providecommand{\BIBforeignlanguage}[2]{{%
\expandafter\ifx\csname l@#1\endcsname\relax
\typeout{** WARNING: IEEEtran.bst: No hyphenation pattern has been}%
\typeout{** loaded for the language `#1'. Using the pattern for}%
\typeout{** the default language instead.}%
\else
\language=\csname l@#1\endcsname
\fi
#2}}
\providecommand{\BIBdecl}{\relax}
\BIBdecl

\bibitem{eventSurvey}
G.~{Gallego}, T.~{Delbruck}, G.~M. {Orchard}, C.~{Bartolozzi}, B.~{Taba},
  A.~{Censi}, S.~{Leutenegger}, A.~{Davison}, J.~{Conradt}, K.~{Daniilidis},
  and D.~{Scaramuzza}, ``Event-based vision: A survey,'' \emph{IEEE
  Transactions on Pattern Analysis and Machine Intelligence}, pp. 1--1, 2020.

\bibitem{cannici2019}
M.~{Cannici}, M.~{Ciccone}, A.~{Romanoni}, and M.~{Matteucci}, ``Asynchronous
  convolutional networks for object detection in neuromorphic cameras,'' in
  \emph{2019 IEEE/CVF Conference on Computer Vision and Pattern Recognition
  Workshops (CVPRW)}, 2019, pp. 1656--1665.

\bibitem{Pantho2021}
\BIBentryALTinterwordspacing
M.~J.~H. Pantho, P.~Bhowmik, and C.~Bobda, ``Towards an efficient cnn inference
  architecture enabling in-sensor processing,'' \emph{Sensors}, vol.~21, no.~6,
  2021. [Online]. Available: \url{https://www.mdpi.com/1424-8220/21/6/1955}
\BIBentrySTDinterwordspacing

\bibitem{Gehrig2018}
\BIBentryALTinterwordspacing
D.~Gehrig, H.~Rebecq, G.~Gallego, and D.~Scaramuzza, ``Asynchronous,
  photometric feature tracking using events and frames,'' \emph{CoRR}, vol.
  abs/1807.09713, 2018. [Online]. Available:
  \url{http://arxiv.org/abs/1807.09713}
\BIBentrySTDinterwordspacing

\bibitem{esim2018}
H.~Rebecq, D.~Gehrig, and D.~Scaramuzza, ``Esim: an open event camera
  simulator,'' in \emph{Proceedings of The 2nd Conference on Robot Learning},
  ser. Proceedings of Machine Learning Research, A.~Billard, A.~Dragan,
  J.~Peters, and J.~Morimoto, Eds., vol.~87.\hskip 1em plus 0.5em minus
  0.4em\relax PMLR, 29--31 Oct 2018, pp. 969--982.

\bibitem{Reichel2015}
P.~Reichel, C.~Hoppe, J.~D\"{o}ge, and N.~Peter, ``Simulation environment for a
  vision-system-on-chip with integrated processing,'' in \emph{Proceedings of
  the 9th International Conference on Distributed Smart Cameras}, ser. ICDSC
  '15.\hskip 1em plus 0.5em minus 0.4em\relax New York, NY, USA: Association
  for Computing Machinery, 2015, p. 20–25.

\bibitem{Chen2020}
G.~{Chen}, H.~{Cao}, J.~{Conradt}, H.~{Tang}, F.~{Rohrbein}, and A.~{Knoll},
  ``Event-based neuromorphic vision for autonomous driving: A paradigm shift
  for bio-inspired visual sensing and perception,'' \emph{IEEE Signal
  Processing Magazine}, vol.~37, no.~4, pp. 34--49, 2020.

\bibitem{sakakibara2018}
M.~Sakakibara \emph{et~al.}, ``A back-illuminated global-shutter cmos image
  sensor with pixel-parallel 14b subthreshold adc,'' in \emph{2018
  ISSCC}.\hskip 1em plus 0.5em minus 0.4em\relax IEEE, 2018, pp. 80--82.

\bibitem{Bi2017}
Y.~{Bi} and Y.~{Andreopoulos}, ``Pix2nvs: Parameterized conversion of
  pixel-domain video frames to neuromorphic vision streams,'' in \emph{2017
  IEEE International Conference on Image Processing (ICIP)}, 2017, pp.
  1990--1994.

\bibitem{pyDVS2016}
G.~P. {García}, P.~{Camilleri}, {Qian Liu}, and S.~{Furber}, ``pydvs: An
  extensible, real-time dynamic vision sensor emulator using off-the-shelf
  hardware,'' in \emph{2016 IEEE Symposium Series on Computational Intelligence
  (SSCI)}, 2016, pp. 1--7.

\bibitem{Katz2012}
M.~L. {Katz}, K.~{Nikolic}, and T.~{Delbruck}, ``Live demonstration:
  Behavioural emulation of event-based vision sensors,'' in \emph{2012 IEEE
  International Symposium on Circuits and Systems (ISCAS)}, 2012, pp. 736--740.

\bibitem{mueggler2017}
\BIBentryALTinterwordspacing
E.~Mueggler, H.~Rebecq, G.~Gallego, T.~Delbruck, and D.~Scaramuzza, ``The
  event-camera dataset and simulator: Event-based data for pose estimation,
  visual odometry, and slam,'' \emph{The International Journal of Robotics
  Research}, vol.~36, no.~2, pp. 142--149, 2017. [Online]. Available:
  \url{https://doi.org/10.1177/0278364917691115}
\BIBentrySTDinterwordspacing

\bibitem{InteriorNet18}
W.~Li, S.~Saeedi, J.~McCormac, R.~Clark, D.~Tzoumanikas, Q.~Ye, Y.~Huang,
  R.~Tang, and S.~Leutenegger, ``Interiornet: Mega-scale multi-sensor
  photo-realistic indoor scenes dataset,'' in \emph{British Machine Vision
  Conference (BMVC)}, 2018.

\bibitem{Whitney2020}
W.~Mason, ``{New frontiers in imaging at DARPA MTO (Conference
  Presentation)},'' in \emph{Infrared Technology and Applications XLVI}, B.~F.
  Andresen, G.~F. Fulop, J.~L. Miller, and L.~Zheng, Eds., vol. 11407,
  International Society for Optics and Photonics.\hskip 1em plus 0.5em minus
  0.4em\relax SPIE, 2020.

\bibitem{pantho2020}
M.~J. {Hossain Pantho}, P.~{Bhowmik}, and C.~{Bobda}, ``Near-sensor inference
  architecture with region aware processing,'' in \emph{2020 IEEE 38th
  International Conference on Computer Design (ICCD)}, 2020, pp. 271--278.

\bibitem{scamp52018}
J.~Chen \emph{et~al.}, ``Scamp5d vision system and development framework,'' in
  \emph{Proceedings of the 12th International Conference on Distributed Smart
  Cameras}, ser. ICDSC ’18.\hskip 1em plus 0.5em minus 0.4em\relax New York,
  NY, USA: Association for Computing Machinery, 2018.

\bibitem{Pang_2019}
\BIBentryALTinterwordspacing
J.~Pang, C.~Li, J.~Shi, Z.~Xu, and H.~Feng, ``$\mathcal{R}^2$ -cnn: Fast tiny
  object detection in large-scale remote sensing images,'' \emph{IEEE
  Transactions on Geoscience and Remote Sensing}, vol.~57, no.~8, p.
  5512–5524, Aug 2019. [Online]. Available:
  \url{http://dx.doi.org/10.1109/TGRS.2019.2899955}
\BIBentrySTDinterwordspacing

\bibitem{Chen2016}
Y.~{Chen}, T.~{Krishna}, J.~{Emer}, and V.~{Sze}, ``14.5 eyeriss: An
  energy-efficient reconfigurable accelerator for deep convolutional neural
  networks,'' in \emph{2016 IEEE International Solid-State Circuits Conference
  (ISSCC)}, 2016.

\bibitem{MOT16}
\BIBentryALTinterwordspacing
A.~Milan, L.~Leal-Taix\'{e}, I.~Reid, S.~Roth, and K.~Schindler, ``{MOT}16: {A}
  benchmark for multi-object tracking,'' \emph{arXiv:1603.00831 [cs]}, 2016,
  arXiv: 1603.00831. [Online]. Available: \url{http://arxiv.org/abs/1603.00831}
\BIBentrySTDinterwordspacing

\bibitem{OpenImages}
A.~Kuznetsova, H.~Rom, N.~Alldrin, J.~Uijlings, I.~Krasin, J.~Pont-Tuset,
  S.~Kamali, S.~Popov, M.~Malloci, A.~Kolesnikov, T.~Duerig, and V.~Ferrari,
  ``The open images dataset v4: Unified image classification, object detection,
  and visual relationship detection at scale,'' \emph{IJCV}, 2020.

\bibitem{mosquito}
\BIBentryALTinterwordspacing
R.~P. M. A. K. P.~P. Chumchu, ``Image dataset of aedes and culex mosquito
  species,'' 2020. [Online]. Available:
  \url{https://dx.doi.org/10.21227/m05g-mq78}
\BIBentrySTDinterwordspacing

\bibitem{Kaiser2016}
J.~{Kaiser}, J.~C. {Vasquez Tieck}, C.~{Hubschneider}, P.~{Wolf}, M.~{Weber},
  M.~{Hoff}, A.~{Friedrich}, K.~{Wojtasik}, A.~{Roennau}, R.~{Kohlhaas},
  R.~{Dillmann}, and J.~M. {Zöllner}, ``Towards a framework for end-to-end
  control of a simulated vehicle with spiking neural networks,'' in \emph{2016
  IEEE International Conference on Simulation, Modeling, and Programming for
  Autonomous Robots (SIMPAR)}, 2016, pp. 127--134.

\end{thebibliography}

\end{document}